\documentclass[%
reprint,%
 amssymb, amsmath,%
 aip,apl,%
groupedaddress,%
]{revtex4-1}
\usepackage[dvipdfmx]{graphicx}
\usepackage[italicdiff]{physics}
\usepackage{upgreek}
\usepackage[colorlinks=true,linkcolor=blue]{hyperref}%
\usepackage{comment}
\expandafter\ifx\csname package@font\endcsname\relax\else
 \expandafter\expandafter
 \expandafter\usepackage
 \expandafter\expandafter
 \expandafter{\csname package@font\endcsname}%
\fi
\begin{document}

\title{Maskless laser processing of graphene}

\author{Fujio Wakaya}%
\email{wakaya@stec.es.osaka-u.ac.jp}
\affiliation{%
Center for Science and Technology at Extreme Conditions, 
Graduate School of Engineering Science, 
Osaka University
1-3 Machikaneyama, Toyonaka, Osaka 560-8531, Japan%
}
\author{Tadashi Kurihara}
\affiliation{%
Center for Science and Technology at Extreme Conditions, 
Graduate School of Engineering Science, 
Osaka University
1-3 Machikaneyama, Toyonaka, Osaka 560-8531, Japan%
}
\author{Nariaki Yurugi}
\affiliation{%
Center for Science and Technology at Extreme Conditions, 
Graduate School of Engineering Science, 
Osaka University
1-3 Machikaneyama, Toyonaka, Osaka 560-8531, Japan%
}
\author{Satoshi Abo}
\affiliation{%
Center for Science and Technology at Extreme Conditions, 
Graduate School of Engineering Science, 
Osaka University
1-3 Machikaneyama, Toyonaka, Osaka 560-8531, Japan%
}
\author{Masayuki Abe}
\affiliation{%
Center for Science and Technology at Extreme Conditions, 
Graduate School of Engineering Science, 
Osaka University
1-3 Machikaneyama, Toyonaka, Osaka 560-8531, Japan%
}
\author{Mikio Takai}
\affiliation{%
Center for Science and Technology at Extreme Conditions, 
Graduate School of Engineering Science, 
Osaka University
1-3 Machikaneyama, Toyonaka, Osaka 560-8531, Japan%
}

\date{Oct. 24, 2014 submitted to Microelectronic Eng.;   
revised on Mar.10, 2015;  accepted on Mar. 25, 2015}

\begin{abstract}
Graphene on a SiO$_2$/Si substrate was removed by ultraviolet pulsed laser irradiation.  Threshold laser power density to remove graphene depended on the graphene thickness.  The mechanism is discussed using kinetic energy of thermal expansion of the substrate surface.  Utilizing the thickness dependence, thickness (or layer-number) selective process for graphene is demonstrated.  Maskless patterning of graphene using laser irradiation in the air is also demonstrated. \\

{\small
Journal reference: Microelectronic Eng. \textbf{141}, 203-206 (2015) 
(published online Apr. 2, 2015)\\
DOI: 10.1016/j.mee.2015.03.049
}
\end{abstract}

\maketitle

\section{Introduction}

Graphene has been drawing attention due to the outstanding electrical and optical characteristics since the so-called scotch-tape method was used to transfer graphene 
from graphite.\cite{Novoselov2004}  
Although extremely high room-temperature carrier mobility in graphene may be utilized for a high-speed transistor, realizing such transistors is still a challenging issue due to such problems as band-gap control \cite{McCann2006,Ohta2006} and mobility suppression by 
extrinsic scatterings.\cite{Bolotin2008,Bolotin2008a}  
Another most realistic application of graphene is a transparent electrode utilizing high transmittance and high electrical conductance of graphene.  Graphene transparent electrodes have been reported in solar cells, touch screen panels, and flat panel 
displays,\cite{Wu2008,Kim2009a,Li2009,Bae2010} which means that mass-production technology is one of the most important issues in this field presently. 

Indium tin oxide (ITO) is widely used for the material of the transparent electrode.  For patterning of ITO, maskless laser process using infrared or ultraviolet (UV) lasers was reported as a simple and fast process comparing to the conventional lithography process with a resist and 
masks.\cite{Takai1994,Yavas1998}  
The mechanism for the ITO etching by laser irradiation was reported to be high-temperature effects, i.e. melting, evaporation, and ablation.\cite{Takai1994,Yavas1998}  

In the previous study,\cite{Wakaya2013} we reported that UV pulsed laser irradiation with a wavelength of 248 nm removed graphene from a SiO$_2$/Si surface if the laser power density is over a threshold value, and that no thickness changes and no defect generations were observed below the threshold power density.  Mechanism for the observed removal was supposed to be mechanical ejection from the substrate surface.  Moreover, a possibility of selective removal of thick graphene was pointed out.  

In this paper, graphene thickness dependence of the threshold laser power density is discussed, and the thickness-selective process using UV pulsed laser irradiation is demonstrated.  Such a selective process should be quite useful because a specific-layer-number graphene is preferable for many applications.  Furthermore, the maskless patterning of the graphene is also demonstrated. 

\section{Experimental}

For the experiments of determining the threshold laser power density to remove graphene from the substrate surface, a Si substrate covered with 100-nm-thick SiO$_2$ layer was used.  Graphene pieces were transferred onto the substrate surface from natural graphite using the scotch-tape process.  The samples after the scotch-tape process were first observed by optical microscopy.  The graphene thickness was determined by atomic-force-microscopy (AFM) observation and Raman spectroscopy with an incident laser wavelength of 488 nm.  The samples were irradiated by KrF excimer laser with a wavelength and a pulse width of 248 nm and 20 ns, respectively.  The laser beam was homogenized and shaped to a stripe (0.4 $\times$ 60 mm$^2$).  The repetition frequency of the pulsed laser was 30 Hz.  The sample stage was moved at 0.6 mm/s to the direction of the 0.4-mm width of the stripe, resulting in 95\% overlap between the successive two pulses and 20 pulse irradiations at every point on the sample.  This overlap and the many irradiations are to avoid the influence of ununiform intensity in the stripe of the laser light.  Accumulation of the laser irradiation effects cannot be expected because the laser pulse width of 20 ns is much shorter than the repetition cycle of 33 ms.  After laser irradiation, the samples were again observed by optical microscopy.  This procedure was repeated with higher laser power density until the graphene was removed.  The laser power density at which the graphene was removed was defined as the threshold laser power density.  The threshold laser power densities for many graphene pieces with various thicknesses were determined to obtain thickness dependence of the threshold power density. 

For the demonstration of maskless patterning using laser irradiation, almost all setups were the same as described above except for the starting material and the sample stage control.  Commercially available graphene-on-SiO$_2$(100nm)/Si samples were used for this purpose.  Almost all regions of the sample surface were covered by single-layer graphene, which was grown by the chemical vapor deposition and transferred onto the SiO$_2$/Si substrate.  The sample stage was not moved during the demonstration.  A 0.4-mm-wide stripe pattern was, therefore, expected to appear by the demonstration. 

\section{Results and discussion}

\begin{figure}[t]
\centering
\includegraphics[width=5.5cm]{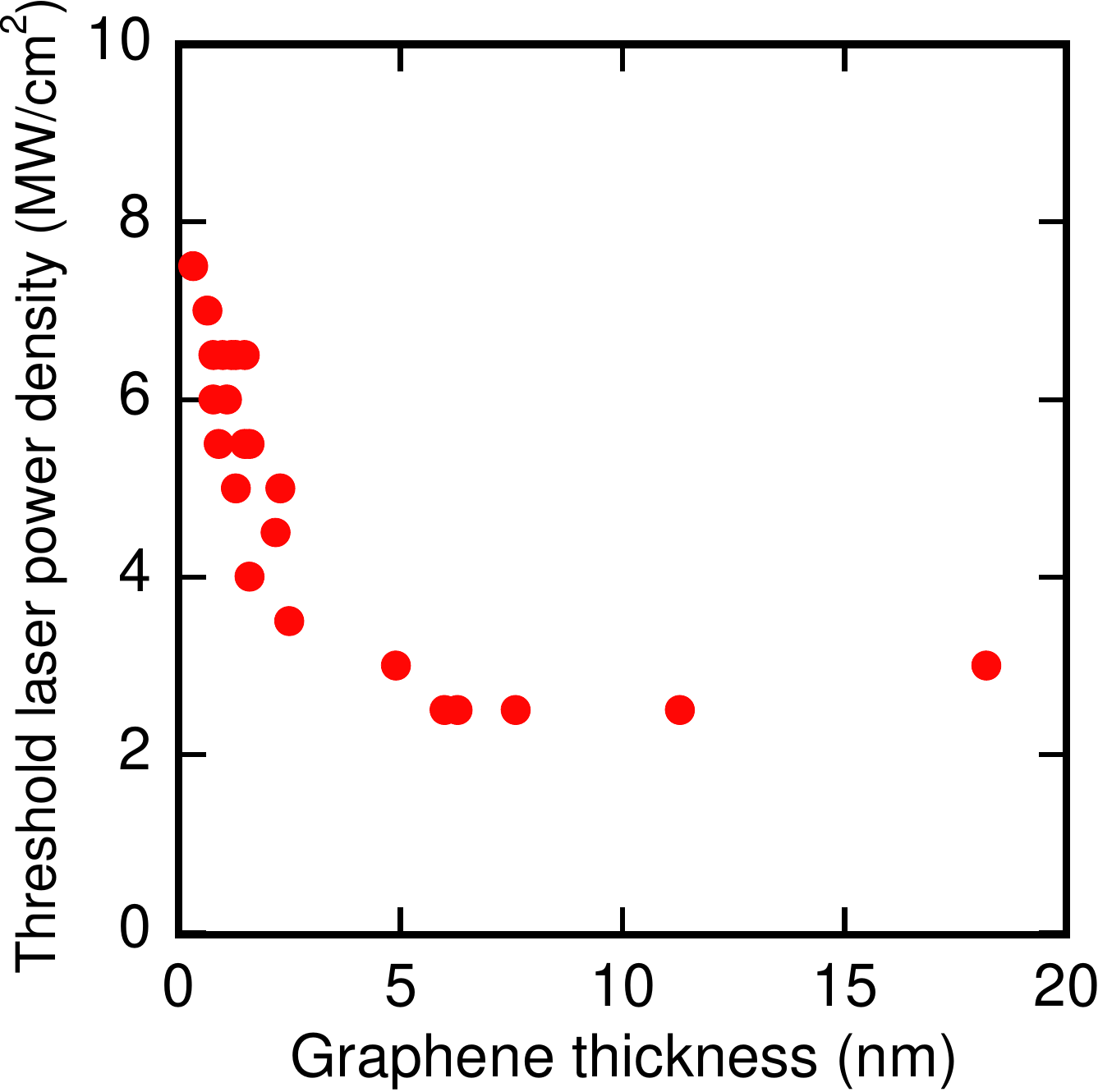}
\caption{
Graphene thickness dependence of threshold laser power density to remove graphene from SiO$_2$/Si substrate.  Thinner graphene needs more laser power.
}
\label{fig:Pth} 
\end{figure}

\begin{figure}[t]
\centering
\vspace*{5mm}
\includegraphics[width=5.5cm]{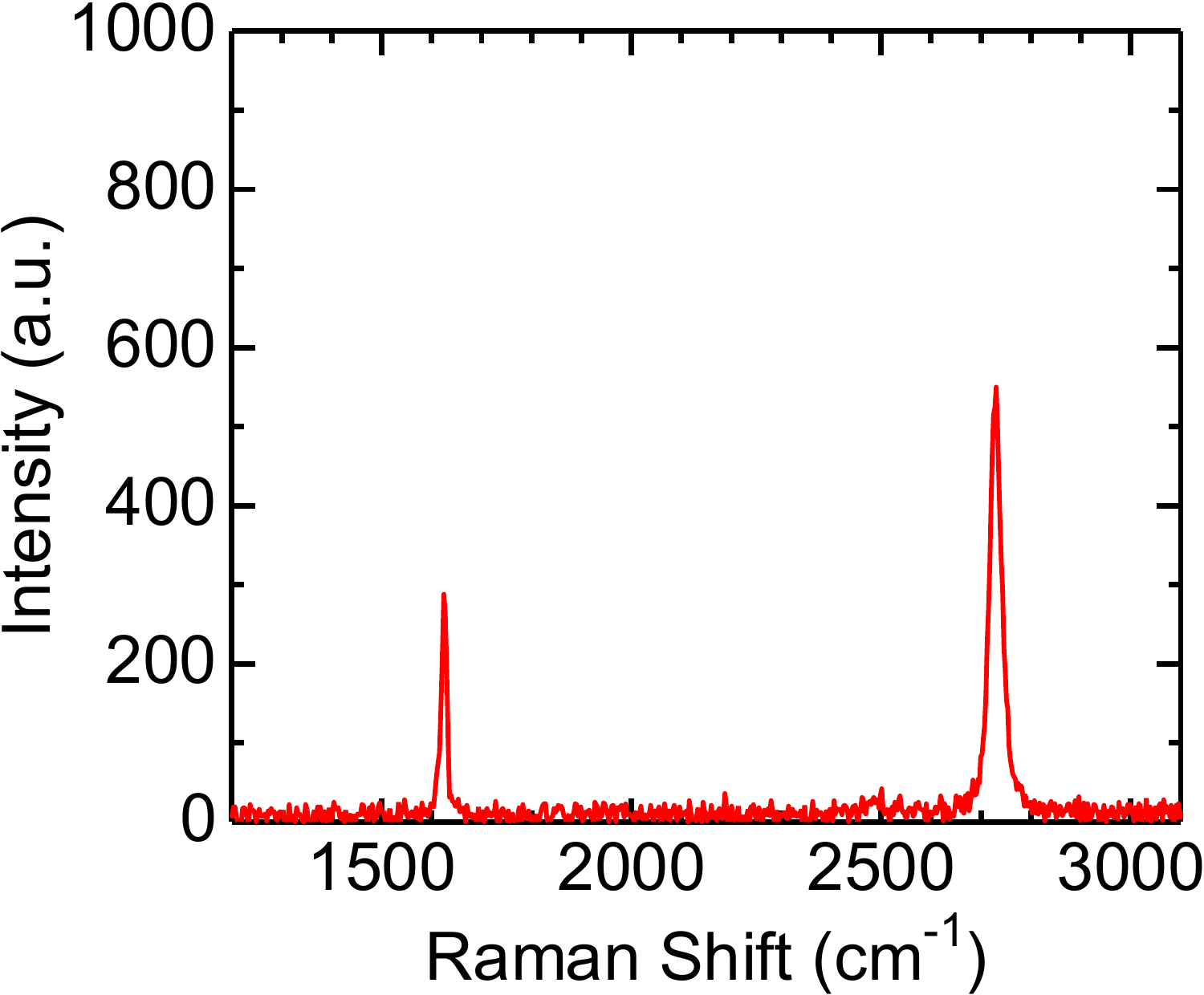}
\caption{
Raman spectrum of graphene.  The graphene is identified as single layer because the ratio of the integrated intensities of the G and D$'$ peaks is 0.21.
}
\label{fig:2} 
\end{figure}

Fig.~\ref{fig:Pth} shows experimentally obtained graphene thickness dependence of the threshold laser power density for removing graphene from the substrate surface.  
It is shown that the thinner graphene needs the higher laser power.  While the AFM measurement was used to determine the graphene thickness, Raman spectroscopy was also used for very thin graphene to be regarded as single-layer graphene.  
Fig.~\ref{fig:2} shows the Raman spectrum for the graphene identified as 
single-layer.\cite{Graf2007}  
As discussed in the dry laser cleaning process,\cite{Arnold2002a} during the laser irradiation the substrate material is thermally expanded with a surface velocity, $v$, of 
\begin{equation}
v=\frac{1+\sigma}{3(1-\sigma)} \frac{\beta I}{c\rho}
\label{eq:velocity}
\end{equation}
Here, $\sigma$ is the Poisson ratio, $\beta$ is the thermal expansion coefficient, $c$ is the specific heat, $\rho$ is the density, and $I$ is the absorbed laser power density.  Ignoring reflection of the laser light and absorption by the graphene, we can estimate the velocity $v$ to be 22 cm/s for 8.0 MW/cm$^2$ laser irradiation.  
For the estimation, $\sigma=0.27$, $\beta=7.2 \times 10^{-6}$ K$^{-1}$, $c =0.72$ J/gK, and 
$\rho= 2.3$ g/cm$^3$ are used for Si at 300 K \cite{Arnold2002a} 
although these parameters should depend on the temperature, and SiO$_2$ is ignored because it is transparent for the 248-nm-wavelength light.  
The kinetic energy per unit area which the graphene could receive from the surface expansion can be estimated as $(1/2)\rho_\mathrm{s} v^2$, where $\rho_\mathrm{s}$ is the sheet density of graphene.  
This energy is $1.8 \times 10^{-8}$ J/m$^2$ for single-layer graphene for 8.0 MW/cm$^2$ laser irradiation.  The energy $(1/2)\rho_\mathrm{s} v^2$ is proportional to the sheet density of graphene, which means that the thicker graphene can receive more energy from the substrate expansion.  We believe that receiving this kinetic energy is the principal mechanism for the observed thickness dependence of the threshold laser power density.  

The adhesion potential between graphene and a SiO$_2$ surface was reported 
as $\sim 10^{-1}$ J/m$^2$,\cite{Koenig2011,Zong2010} 
which is measured at room temperature and seven orders of magnitude larger than the kinetic energy discussed above.  The large discrepancy may be due to the temperature rise during the laser irradiation.  When a laser light with a power density of I is absorbed by a thick Si substrate, the temperature T at the Si surface after 
time t is \cite{Toyoda1990}
\begin{equation}
T=\frac{2 I }{k} \sqrt{\frac{Dt}{\pi}}
\label{eq:temp}
\end{equation}
where $k$ and $D$ are the thermal conductivity and thermal diffusivity of Si, respectively.  
Although the layer structure in the present work is different from the thick Si substrate without any other layers, the temperature at the graphene and the SiO$_2$ layer can 
be estimated using Eq.~(\ref{eq:temp}) 
because the absorption at the thin graphene and the SiO$_2$ layer can be ignored for a rough estimation.  The temperature after 20 ns from 8.0 MW/cm$^2$ irradiations is actually estimated to be 790 $^\mathrm{o}$C.  
For the estimation, $k = 168$ W/mK \cite{Rikanempyou1990} and $D = k/c\rho$ are used.  
The thermal vibration energy per each atom can be estimated as $\sim k_\mathrm{B}T$, 
which is $1.5 \times 10^{-20}$ J for 790 $^\mathrm{o}$C.  
The atom density of graphene is $3.8 \times 10^{19}$ m$^{-2}$.  
The thermal vibration energy per unit area is, therefore, 
$\sim 5.7 \times 10^{-1}$ J/m$^2$, 
which is on the same order as the room-temperature adhesion potential between 
graphene and SiO$_2$.\cite{Koenig2011,Zong2010}  
The estimated high temperature, therefore, might affect the adhesion potential effectively through the thermal vibration of the atoms.  The absorption coefficient of the 248-nm-wavelength light for the graphite along the c axis and the corresponding absorption length are 0.13 nm$^{-1}$ and 
7.7 nm, respectively,\cite{Borghesi1991a} 
which means that relatively large absorption occurs at the graphene layer especially if the graphene is thick comparing to the absorption length.  A numerical simulation taking the reflection at the top surface and the interface, the absorption at the graphene layer, the heat conduction at all layers and the thermal expansion of SiO$_2$ and Si layers into account is necessary to discuss the thermal expansion and the temperature in more detail.  

\begin{figure}[t]
\centering
\includegraphics[width=5cm]{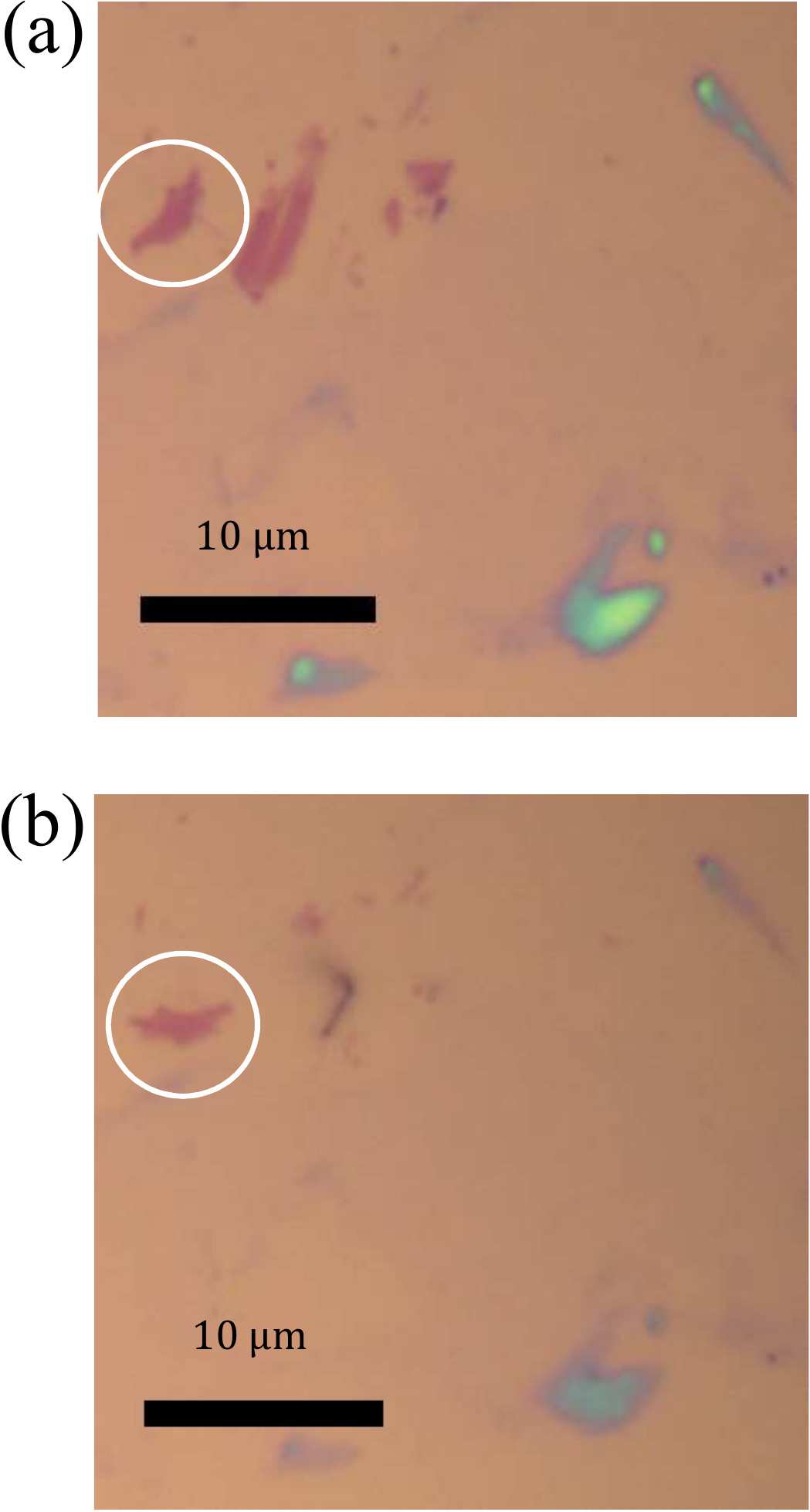}
\caption{
Optical microscope images of sample surface (a) after scotch-tape process, 
(b) after 3.0 MW/cm$^2$ laser irradiation.  The 4.1-nm-thick graphene indicated by the white circle slightly changes the position and apparently changes the rotation angle.  
}
\label{fig:3} 
\end{figure}
The authors in Ref.~\onlinecite{Dhar2011} also reported that the threshold laser energy depended on the graphene thickness, which is quite similar to the result in the present work.  
Their interpretation is, however, different from that in the present work.  They argued that the mechanism for the graphene removal was thermal ablation and attributed the observed graphene thickness dependence to that the specific heat depends on the graphene thickness.  They, however, observed that a part of the graphene edge was folded onto the graphene after laser irradiation.  As a similar example, we observed that a graphene piece was found at a different position with a different rotation angle after laser irradiation as shown 
in Fig.~\ref{fig:3}.
Such experimental observation suggests that the mechanism for the graphene removal is mechanical ejection from the surface.  The large discrepancy of the estimated kinetic energy to the adhesion energy reported in the literature, however, may suggest that both of the mechanical and thermal effects should be considered. 

\begin{figure}[t]
\centering
\includegraphics[width=5cm]{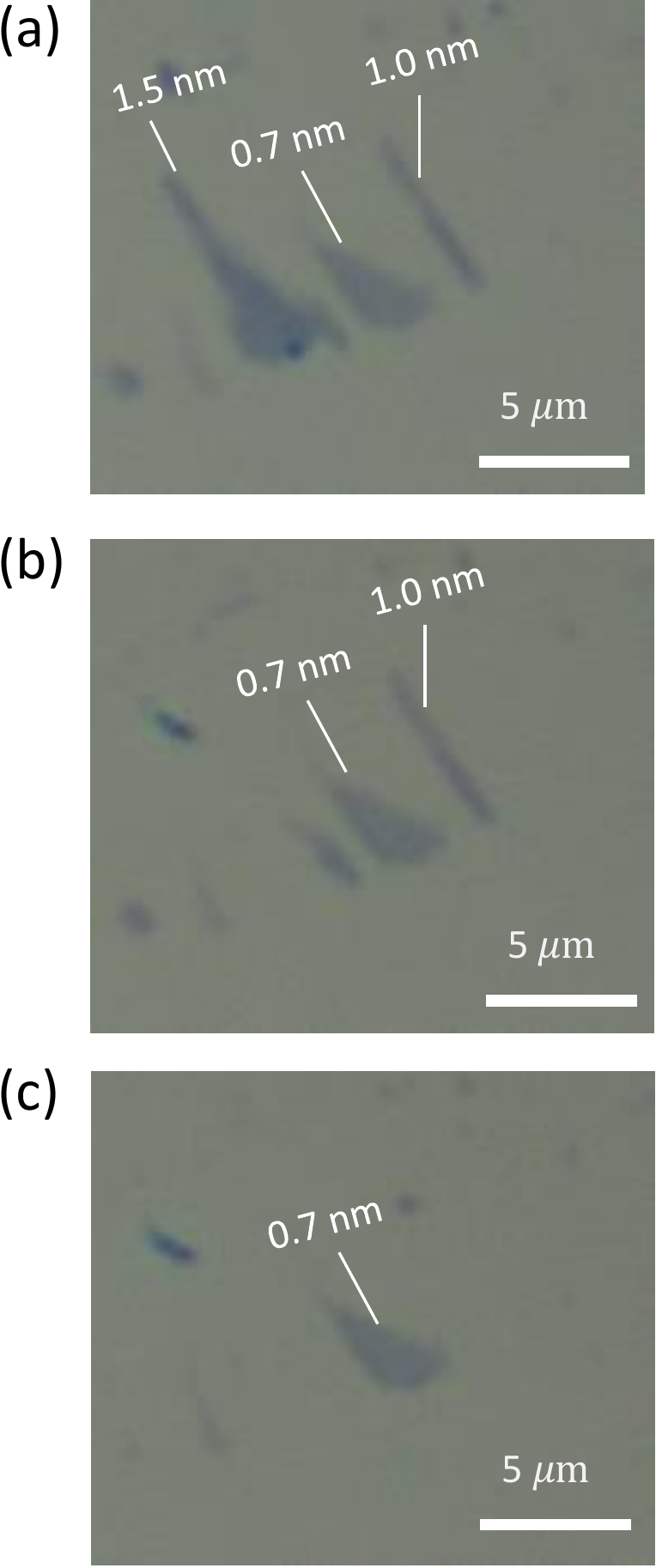}
\caption{
Optical microscope images of sample surface (a) after scotch-tape process, (b) after 5.5 MW/cm$^2$ laser irradiation, and (c) after 6.5 MW/cm$^2$ laser irradiation.  The graphene thicknesses are indicated by white letters in the images.  The thickness (or layer-number) selective process is demonstrated. 
}
\label{fig:layer-selective} 
\end{figure}

\begin{figure}[t]
\centering
\includegraphics[width=6cm]{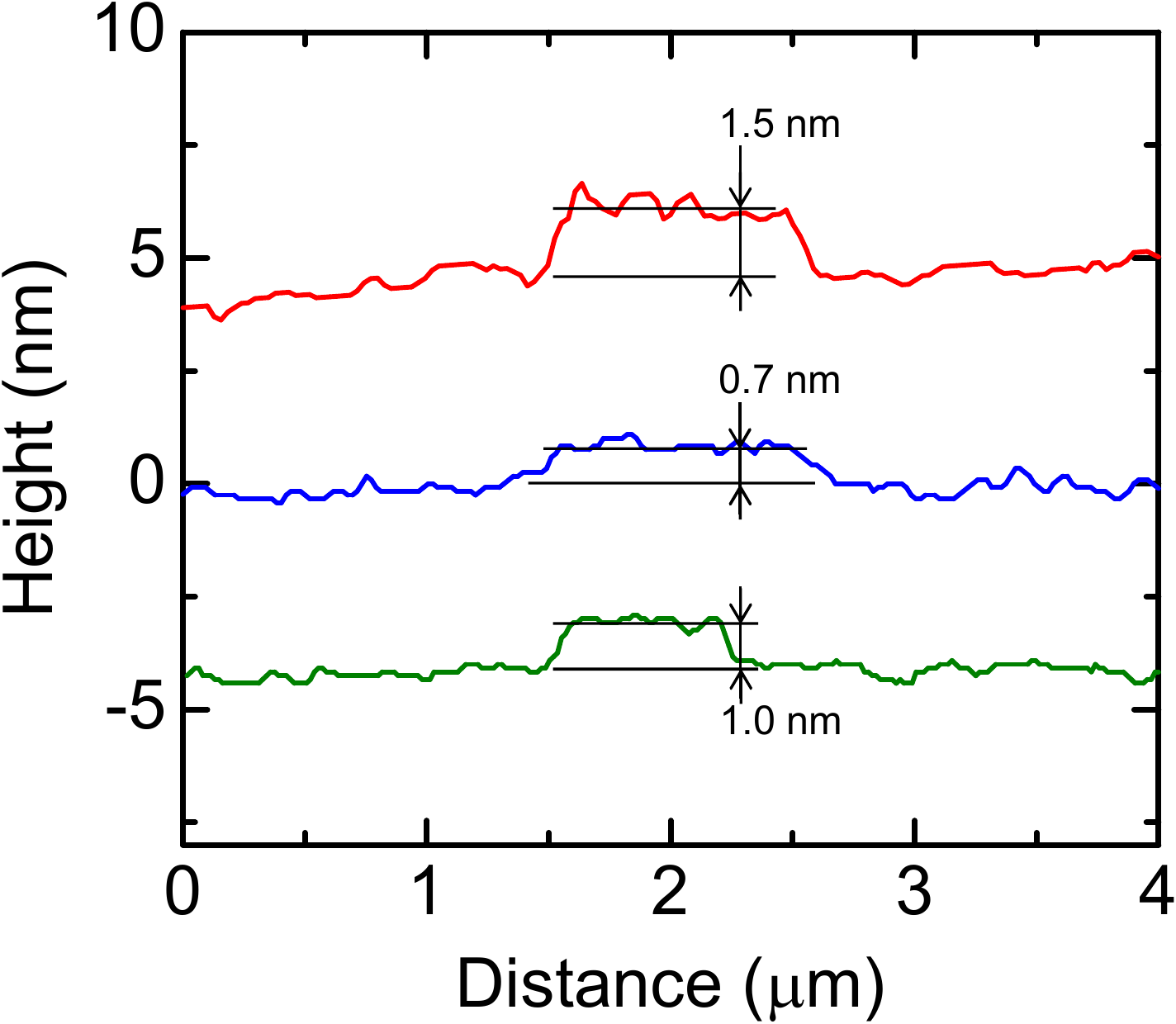}
\caption{
AFM cross-sectional plots of graphenes shown in Fig.~\ref{fig:layer-selective}.
}
\label{fig:5} 
\end{figure}

The graphene thickness dependence of the threshold power density shown in 
Fig.~\ref{fig:Pth}  suggests a possibility of thickness (or layer-number) selective process.  
Fig.~\ref{fig:layer-selective} shows a demonstration of such a process.  
There are three graphene pieces with a thickness of 0.7 nm, 1.0 nm, and 1.5 nm as shown 
in Fig.~\ref{fig:layer-selective}(a).  The AFM cross-sectional plots of the graphene pieces 
are shown in Fig.~\ref{fig:5}.  
After 5.5 MW/cm$^2$ laser irradiation, only the 1.5-nm-thick graphene was removed as shown in 
Fig.~\ref{fig:layer-selective}(b).  
The 1.0-nm-thick graphene was removed by the 6.5 MW/cm$^2$ laser irradiation as shown 
in Fig.~\ref{fig:layer-selective}(c).  
As a result, only the thinnest graphene with a thickness of 0.7 nm remains on the substrate.  This experiment successfully demonstrates the thickness selective process of the graphene by UV pulsed laser irradiation. 

\begin{figure}[t]
\centering
\vspace*{5mm}
\includegraphics[width=5cm]{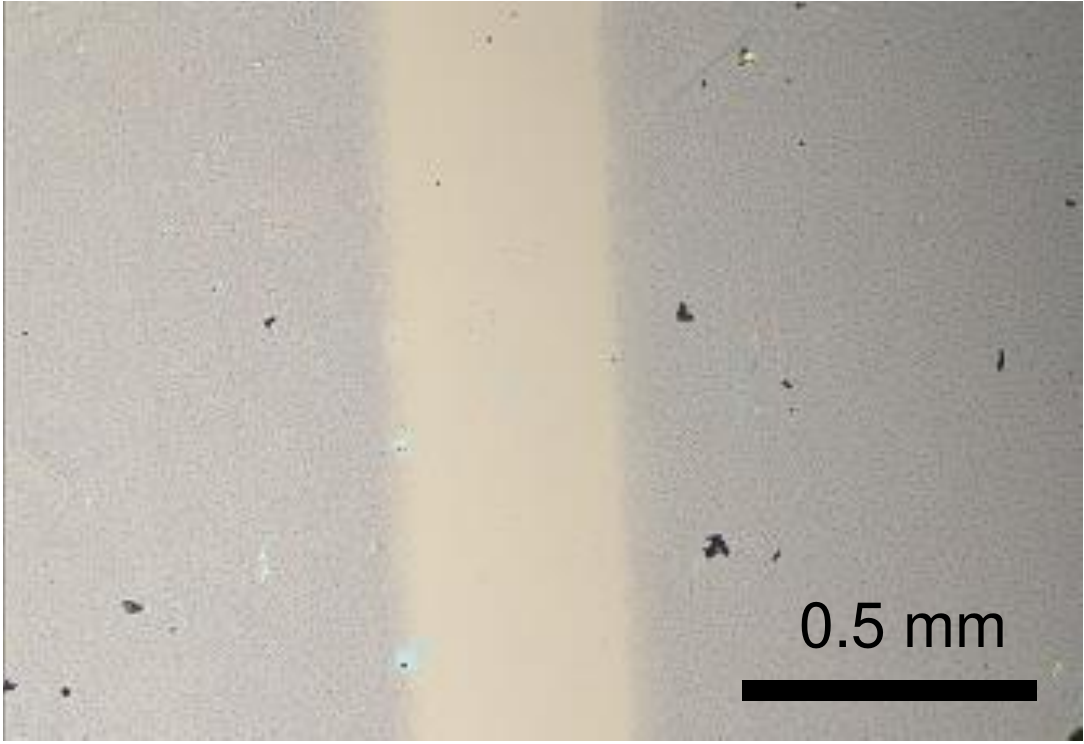}
\caption{
Optical microscope image of sample surface after 10 MW/cm$^2$ laser irradiation.  Graphene only in the irradiated region, which is the central part of the image, is removed.  Maskless patterning of graphene using laser irradiation in the air is demonstrated.  
}
\label{fig:patterning} 
\end{figure}

A demonstration result of the maskless patterning of graphene in the air by laser irradiation is shown in Fig.~\ref{fig:patterning}.   
The central part of the single-layer graphene/SiO$_2$/Si sample surface was irradiated by stripe-shaped laser light at 10 MW/cm$^2$.  
The 0.4-mm-wide stripe where no graphene existed was clearly observed in the figure.  If the laser light is shaped as a spot and the sample stage is computer-controlled, an arbitrary pattern can be realized by laser irradiation in the air without masks, which should be useful in the mass-production process.  The graphene material used for the demonstration was not a huge single crystal but consisted of many pieces of grain, the size of which was typically several $\upmu$m.  Possible mechanism for the demonstration is, therefore, that each piece of the graphene is ejected without breaking the covalent bond between the carbon atoms when the substrate is irradiated by the laser.

\section{Summary}

Graphene pieces on a SiO$_2$/Si substrate were removed by UV pulsed laser irradiation.  
The threshold power density to remove graphene depended on the graphene thickness.  The mechanism was proposed using the substrate thermal expansion as is well known in the dry laser cleaning process.  

Utilizing the thickness dependence of the threshold laser power density, thickness selective process for graphene was demonstrated.  The thickness selective process, or layer-number selective process, is quite interesting because a specific layer-number graphene is preferable in many applications.  

Maskless patterning of graphene using laser irradiation in the air was demonstrated using a 
SiO$_2$/Si substrate covered with single-layer graphene.  This process will contribute to the mass production of graphene devices.

\begin{acknowledgments}
The authors acknowledge financial support from the Japan Society for Promotion of Science with KAKENHI Grant Number 25420287.
\end{acknowledgments}

\section*{References}


%

\end{document}